\documentclass{emulateapj}
\usepackage{amsmath}
\usepackage{graphicx}
\shorttitle{Perpendicular Current-Driven Instability}
\shortauthors{Riquelme \& Spitkovsky}

\begin{document}

\title{Magnetic Amplification by Magnetized Cosmic Rays in SNR Shocks}

\author{Mario A. Riquelme and Anatoly Spitkovsky}
\affil{Department of Astrophysical Sciences, Princeton University, Princeton, NJ 08544}
\email{marh@astro.princeton.edu, anatoly@astro.princeton.edu}

\begin{abstract}
X-ray observations of synchrotron rims in supernova remnant (SNR) shocks show evidence of efficient electron acceleration and strong magnetic field amplification (a factor of $\sim 100$ between the upstream and downstream medium). This amplification may be due to plasma instabilities driven by shock-accelerated particles, or cosmic rays (CRs), as they propagate ahead of the shocks. One candidate process is the cosmic ray current-driven (CRCD) instability (Bell 2004), caused by the electric current of ``unmagnetized" CRs (i.e., CRs whose Larmor radii are much larger than the length scale of the CRCD modes) propagating parallel to the upstream magnetic field. Particle-in-cell (PIC) simulations have shown that the back-reaction of the amplified field on CRs would limit the amplification factor of this instability to less than $\sim 10$ in galactic SNRs (not including the additional field compression at the shock). In this paper, we study the possibility of further amplification driven near shocks by ``magnetized" CRs, whose Larmor radii are smaller than the length scale of the field that was previously amplified by the CRCD instability. We find that additional amplification can occur due to a new instability, driven by the CR current perpendicular to the field, which we term the {\it perpendicular current-driven instability} (PCDI). We derive the growth rate of this instability, and, using PIC simulations, study its non-linear evolution. We show that the maximum amplification of PCDI is determined by the disruption of CR current, which happens when CR Larmor radii in the amplified field become comparable to the length scale of the instability. We find that, in regions close to the shock, PCDI grows on scales smaller than the scales of the CRCD instability, and, therefore, it results in larger amplification of the field (amplification factor up to $\sim 45$). One possible observational signature of PCDI is the characteristic dependence of the amplified field on the shock velocity, $B^2 \propto v_{sh}^2$, which contrasts with the one corresponding to the CRCD instability acting alone,  $B^2 \propto v_{sh}^3$. Our results strengthen the idea of CRs driving a significant part of the magnetic field amplification observed in SNR shocks. 

\end{abstract}

\keywords{ISM: magnetic filed  --- cosmic rays --- supernova remnants --- jets and outflows}

\section{Introduction}
\label{sec:intro}
Cosmic rays (CRs) up to $\sim 10^{15} $eV are most likely accelerated in galactic supernova remnant (SNR) shocks via the diffusive shock acceleration (DSA) mechanism \citep{Krymsky77,AxfordEtAl77, Bell78, BlandfordEtAl78}.  This mechanism is believed to be highly non-linear, due to the expected coupling between the evolution of the thermal plasma, the CRs, and the magnetic turbulence in the shock vicinity. Indeed, there is now observational evidence suggesting that electrons are being accelerated in the non-relativistic shocks of SNRs, and that significant magnetic turbulence is being produced as part of the process \citep[see, e.g.][]{Ballet06, UchiyamaEtAl07}. The evidence comes from X-ray observations of SNRs that show the existence of thin, non-thermal rims, which are interpreted as synchrotron emission by TeV electrons accelerated at the shocks. The rapid variability and thinness of the rims (which depend on the synchrotron cooling time of the electrons) have allowed to estimate the strength of the field, suggesting downstream amplitudes $\sim 100$ times larger than typically expected in the ISM of the Galaxy. This implies a significant amplification even accounting for the compression of the field at the shock, which would contribute a factor of $\sim 4$ to the growth of the field. 

The origin of this strong amplification is an open question. On the one hand, using MHD shock simulations, \cite{GiacaloneEtAl07} showed that the presence of density and magnetic fluctuations in the upstream medium of non-relativistic shocks may introduce vorticity into the flow and produce field amplification at the shock itself \citep[see, also,][]{SironiEtAl07}. This mechanism would produce a magnetic enhancement larger than expected from the simple shock compression of the upstream field. On the other hand, it has been proposed that the magnetic field may be amplified by plasma instabilities driven by the CRs themselves, as they propagate through the upstream medium of shocks \citep{Bell04}. This idea is interesting because, besides helping to explain the field amplification, the implied magnetic turbulence would increase the CR confinement to the shock vicinity, enhancing the efficiency of DSA. Due to this, understanding the extent to which magnetic fields are amplified by CRs, as well as characterizing the corresponding magnetic turbulence, is essential for any realistic model of DSA.

The cosmic ray current-driven (CRCD) instability proposed by \cite{Bell04} has gathered considerable attention recently. The CRCD instability consists of circularly polarized Alfv\'{e}n-type waves, driven by the electric current of CRs propagating along the magnetic field lines. It requires the CR Larmor radii, $R_{L,cr}$, to be much larger than the wavelength of the fastest growing mode, $\lambda_{CRCD}= cB_0/J_{cr}$, where $B_0$, $J_{cr}$, and $c$ are the magnitude of the initial magnetic field, the CR current, and the speed of light, respectively. This feature makes this instability different from the previously proposed resonant instability, in which the waves grow at wavelengths comparable to $R_{L,cr}$ \citep{KulsrudEtAl69}. The growth rate of the CRCD instability is given by $\gamma_{CRCD}= J_{cr}(\pi/\rho c^2)^{1/2}$ \citep{Bell04}, where $\rho$ is the mass density of the background plasma. For the typical conditions of CR-modified SNR shocks (where the energy densities of CRs, $u_{cr}$, downstream thermal plasma, $u_{th}$, and magnetic field,$u_B$, satisfy $u_{th} \gtrsim u_{cr} \gg u_B$), this rate is expected to be significantly larger than that of the resonant instability.
 
The non-linear evolution of the CRCD instability has been studied using both MHD \citep{Bell04, ZirakashviliEtAl08} and particle-in-cell (PIC) \citep{NiemiecEtAl08, RiquelmeEtAl09, StromanEtAl09} simulations. Both kinds of studies showed that the CRCD instability has the potential of growing to very non-linear amplitudes as long as the CR current is kept constant, i.e., if the back-reaction on the CR trajectories is neglected. In particular, \cite{RiquelmeEtAl09} found that, at constant CR current, the field growth stops when the Alfv\'{e}n velocity of the background plasma, $v_{A}$, gets close to the CR drift velocity, $v_{d,cr}$. However, PIC studies show that if the CR back-reaction is considered, the field can also saturate due to the trapping of the CRs in the amplified field. When $R_{L,cr}$ becomes comparable to the dominant length scale of the magnetic turbulence, the CRs get strongly deflected, which significantly reduces their current and quenches the magnetic growth. This is the most likely saturation mechanism in the case of SNRs, which implies a {\it maximum} amplification factor of $\sim 10$ in the upstream medium of shocks, and confirms the CRCD instability as a viable mechanism for magnetic turbulence generation in SNRs \citep{RiquelmeEtAl09}. When the field compression at the shock is considered, this amplification may increase by an extra factor of $\sim 4$. However, since the upstream amplification factor of $10$ constitutes an upper limit, the CRCD instability alone would not be enough to account for the factor of $\sim 100$ inferred from the X-ray observations.

In this paper we study the possibility of magnetic amplification beyond the saturation of the CRCD instability. The situation we explore is one where CRs propagate through a plasma where previous  CRCD magnetic turbulence has already been produced on scales larger than the typical CR Larmor radius, $R_{L,cr}$. Thus, our analysis would be applicable to the lower energy CRs, which are more confined to the shock vicinity. This is motivated by the fact that the CRCD instability is probably driven first by the highest energy CRs, in regions far upstream from the shock. This instability saturates when the Larmor radius of these high energy particles is about the length scale of the pre-amplified magnetic fluctuations, $\lambda_{0}$. Thus, as the shock approaches, a large fraction of the CR energy will be carried by ``magnetized" CRs, i.e., CRs with  $R_{L,cr}$ smaller than $\lambda_{0}$. Then, near the shock, CRs will find regions of pre-amplified, quasi-transverse field, which can be considered homogeneous on scales $\sim R_{L,cr}$.

If $R_{L,cr} \lesssim \lambda_{0}$, then CRs will not easily diffuse through the regions of pre-amplified field. Instead, they will protrude into these regions only by a distance of $\sim R_{L,cr}$. This situation will produce a CR current, $\vec{J}_{cr}$, perpendicular to the initial, pre-amplified field, $\vec{B}_0$, due to the coherent deflection experienced by CRs in the ``homogeneous" (on scales of $\sim R_{L,cr}$) magnetic field. 
We propose that, under these conditions, an extra magnetic amplification is possible due to a new instability: the {\it perpendicular current-driven instability} (PCDI), which consists of purely growing, compressional waves that arise when $\vec{J}_{cr}$ is perpendicular to $\vec{B}_0$. 

As we show below, the PCDI would produce larger magnetic amplifications compared to the CRCD instability acting alone. In addition, this instability can amplify magnetic fluctuations on scales comparable to the Larmor radii of the lowest energy CRs, which would improve their diffusion and increase the efficiency of their acceleration at the shock. If this did not happen, the large-scale, transverse fields generated by CRCD instability would hamper the low energy CR diffusion, which would probably shut off the acceleration of the CRs that cause the CRCD amplification in the first place.

In \S \ref{sec:physics}, we explain the physics of the PCDI. First, we show how the penetration of CRs into the regions of pre-amplified field can give rise to a current, $\vec{J}_{cr}$, perpendicular to the pre-amplified field, $\vec{B}_{0}$, and why we expect that situation to happen in the upstream medium of SNR shocks. Then, we show how the presence of a $\vec{J}_{cr}$ perpendicular to $\vec{B}_{0}$ can produce the PCDI, and derive its dispersion relation.  In \S \ref{sec:simulations}, we show the results of our study of the PCDI using PIC simulations. First, we model the non-linear evolution of the instability assuming a constant CR current, i.e., ignoring the back-reaction on the CRs, and provide a simple analytical model for the non-linear behavior of the PCDI modes. Second, we show the results of a series of simulations that model the non-linear evolution of the PCDI including the full CR dynamics. The main focus is to show the role of the CR back-reaction on the final saturation of the instability. In \S \ref{sec:discussion}, we apply our results to the case of SNR shocks. Finally, in \S \ref{sec:conclusions} we summarize our results and present our conclusions. 
 
 \section{Physics of PCDI}
 \label{sec:physics}
The perpendicular current-driven instability (PCDI) is caused by a CR current, $\vec{J}_{cr}$, perpendicular to the ambient magnetic field, $\vec{B}_0$. In \S \ref{sec:anticorrelation}, we show why we expect this $\vec{J}_{cr} \perp \vec{B}_0$ to actually occur in the precursor of SNR shocks. Then, in \S \ref{sec:dispersion} we explain the physics of the PCDI growth, and derive its dispersion relation. 
 
\subsection{Perpendicular Current in SNR Shock Precursors}
\label{sec:anticorrelation}
\begin{figure}
\includegraphics[width=8.6cm]{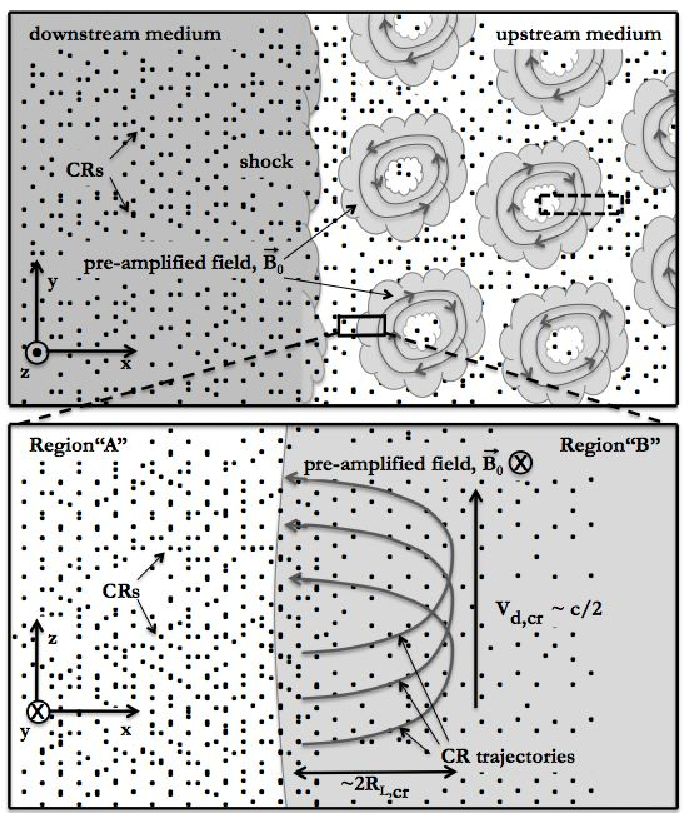}
\caption{A diagram with the global picture of formation of zones with CR current perpendicular to the pre-amplified magnetic field, $\vec{B}_0$, in the upstream medium of SNR shocks. The {\bf upper panel} shows the downstream and upstream regions of a generic shock. Regions of pre-amplified field in the upstream are represented by grey, doughnut-like shapes. Black dots represent ``magnetized" CRs, i.e., CRs whose Larmor radii are smaller than the size of the regions of pre-amplified field. The {\bf lower panel} shows a zoomed and rotated (notice the orientation of coordinate axes in both panels) image of the small, solid-line rectangle shown in the upper panel. This image is divided into region ``A" (with low initial magnetic field), and region ``B" (with the pre-amplified magnetic field $\vec{B}_0$). The CRs, coming from region ``A", can penetrate into region ``B" by a distance $\sim 2R_{L,0}$, which defines the interface into which CRs can protrude. As indicated in the figure, the coherent deflection of the CRs in this region produces a net mean velocity along $\hat{z}$, of magnitude $\sim c/2$. Finally, the small, dashed-line rectangle depicted in the upper panel represents the type of regions that we study in \S \ref{sec:crback} by using two-dimensional PIC simulations.}
 \label{fig:figure2}
 \end{figure}
In this section, we show that a significant CR mean velocity perpendicular to an initial field, $\vec{B}_0$, can arise in the upstream medium of SNR shocks, if there is a significant anti-correlation between the CR number density, $n_{cr}$, and the strength of the magnetic field.

If CRs propagate through a smooth medium, as may be the case shortly after the shock formation, there is no reason to think that a significant anti-correlation between $n_{cr}$ and $|\vec{B_0}|$ can happen. However, if some magnetic turbulence has already been amplified, then this anti-correlation can exist. For instance, numerical MHD and PIC simulations of the CRCD instability (see Fig. 4 of \cite{Bell04} or Figs. 10 and 12 of \cite{RiquelmeEtAl09}) show how this instability produces turbulence characterized by typical plasma density contrast of $\sim 10$, with frozen-in magnetic field concentrated in the high density regions. As discussed in \cite{RiquelmeEtAl09}, this kind of turbulence can be produced far upstream from the shock by high-energy CRs ``escaping" at $\sim c/2$ \citep[see also][]{ZirakashviliEtAl08}. This implies that lower energy CRs, which are more confined to the shock, will encounter the upstream medium with previously amplified magnetic turbulence. 
This is the situation depicted in the upper panel of Figure \ref{fig:figure2}, which shows the regions in the upstream of a shock containing a pre-amplified magnetic field, $\vec{B_0}$ (diagrammatically represented by doughnut-like shapes in the figure; note that, in reality, these ``doughnuts" are not necessarily in one plane) surrounded by CRs (indicated by black dots). Since the saturation of the CRCD instability happens when the Larmor radius of the highest energy CRs is close to the length scale of the pre-amplified field, $\lambda_{0}$, then most CRs near the shock are expected to have Larmor radii, $R_{L,cr}$, that satisfy $R_{L,cr} < \lambda_{0}$. We will refer to CRs satisfying this condition as ``magnetized" CRs.

In this scenario, CRs will not be able to diffuse through the regions of pre-amplified field. Instead, they will only be able to penetrate into these regions by a distance of $\sim 2 R_{L,cr}$. The lower panel in Figure \ref{fig:figure2} depicts this situation. It shows a zoomed image of the interface between a region of low and high upstream magnetic field (rectangular box in the upper panel).This plot is divided into two regions. Region ``A" (left), which contains the CRs and where there is almost no initial magnetic field, and region ``B" (right), where there is a pre-amplified magnetic field, $\vec{B}_0$, which has a length scale, $\lambda_{0}$, larger than the CR Larmor radii, $R_{L,cr}$. Since $R_{L,cr} < \lambda_{0}$, CRs protruding into region {\it B} (by a distance $\sim 2 R_{L,cr}$) will not experience random scatterings due to fluctuations in the field. Instead, they will just complete a fraction (typically $\sim 1/2$) of a Larmor gyration before being scattered back into region {\it A}. The fact that CRs complete only a fraction of a Larmor cycle in region {\it B} implies that, while in this region, they will tend to move parallel to $\vec{\nabla}n_{cr} \times \vec{B}_0$ (i.e., along $\hat{z}$ in Figure \ref{fig:figure2}). Thus,  if CRs have positive electric charge, they will produce a current, $\vec{J}_{cr}$ perpendicular to the pre-amplified field\footnote{Since negative CRs would move in the opposite direction, their negative charge would finally make their current point parallel to the one of positive CRs, implying an even larger $\vec{J}_{cr}$.}. Since CRs are relativistic, their mean velocity perpendicular to $\vec{B}_0$ will be a fraction (of the order of $\sim 1/2$) of the speed of light, larger than the typical drift velocities along the field that drive the CRCD instability. 
This suggests that it is possible to find upstream regions near SNR shocks where $\vec{J}_{cr}$ can be quasi-perpendicular to $\vec{B}_0$. 
The formation of the perpendicular mean CR velocity at $\sim c/2$ 
is confirmed in \S \ref{sec:crback} using two-dimensional PIC simulations.

\subsection{PCDI Dispersion Relation}
\label{sec:dispersion} 
The presence of the CR current, $\vec{J}_{cr}$, perpendicular to the initial field, $\vec{B}_0$, induces a return current, $\vec{J}_{ret}$, almost equal to $-\vec{J}_{cr}$ in the background\footnote{
It is interesting to note that, even though charged particles can not easily move across magnetic field lines, in the case when $\vec{J}_{cr} \perp \vec{B}_0$, $\vec{J}_{ret}$ is still almost equal to $-\vec{J}_{cr}$. In this case the compensating current is due to the polarization drift of particles, $\vec{v}_p = (mc^2/eB_0^2)d\vec{E}/dt$, where $m$ and $e$ are the particle mass and charge, and $\vec{E}$ is the electric field. It is possible to show from the Amp\`{e}res's law that, if $v_{A,0}/c \ll 1$, the return current produced by this drift (acting mainly on the ions) almost completely compensates the CR current (see also Lyutikov 2009).}. Because of this, a constant force, $-\vec{J}_{cr}\times\vec{B}_0$, will push the plasma homogeneously. Thus, if plasma density fluctuations, $\delta \rho$, exist, this $-\vec{J}_{cr}\times\vec{B}_0$ force will accelerate the low density regions more than the ones with higher density. As we will see below, this differential plasma acceleration in the direction perpendicular to $\vec{B}_0$ can amplify the transverse component of the magnetic field, $\delta \vec{B}$, and enhance the density fluctuations in the plasma, producing the exponential growth of the PCDI modes. We note that this instability starts from a non-equilibrium state, where the whole plasma is being accelerated in the direction of $-\vec{J}_{cr}\times\vec{B}_0$.

We derive the dispersion relation of the PCDI in the MHD limit. We assume the existence of an externally imposed CR current, $\vec{J}_{cr} \perp \vec{B}_0$. The way this CR current is produced and the back-reaction of the amplified field, $\vec{B}$ ($=\vec{B}_0 + \delta \vec{B}$), on the CR trajectories is not modeled in this derivation. The current carried by the background plasma is $c\nabla \times \vec{B}/4\pi - \vec{J}_{cr}$, which includes the return current induced in the background.  So the MHD equations will only be modified by an extra term in the force per unit volume on the plasma:
\begin{equation}
\frac{\partial \rho}{\partial t} + \vec{\nabla}\cdot (\rho \vec{v})  =  0
\end{equation}
\begin{equation}
\rho (\frac{\partial \vec{v}}{\partial t} + (\vec{v}\cdot\vec{\nabla})\vec{v}) + \nabla p - 
\frac{1}{4\pi}(\vec{\nabla}\times \vec{B})\times\vec{B} + \frac{\vec{J}_{cr}}{c}\times\vec{B} = 0
\label{eq:euler2} 
\end{equation}
\begin{equation}
\frac{\partial \vec{B}}{\partial t} - \nabla \times (\vec{v} \times \vec{B}) = 0,
\end{equation} 
where $\vec{v}$ is the plasma fluid velocity and $p$ is its pressure. Although the CRs can also include accelerated electrons, which would reduce the CR net charge, in general CRs may carry a charge density, $\rho_{c,cr}$, that must be compensated by the background plasma. This extra plasma charge should introduce an electrostatic term in Equation (\ref{eq:euler2}) due to plasma charge density fluctuations. It is possible to show that this term is second order in perturbation theory, so it can be neglected in our linear analysis. We refer the reader to Appendix \ref{app:B} for further details.

If each physical quantity $\xi$ is separated into an initial, homogeneous component, $\xi_0$, plus a small perturbation, $\delta (\xi \ll \xi_0)$, (so $\vec{B} = \vec{B}_0 + \delta \vec{B}$, $\vec{v} = \vec{v}_0 + \delta \vec{v}$, and $p = p_0 + \delta p$), then the MHD momentum equation will have a zeroth order part:
\begin{equation}
\rho_0\frac{\partial \vec{v}_0}{\partial t} + \vec{J}_{cr} \times \vec{B}_0=0,
\label{eq:euler0}
\end{equation}
which represents homogeneous acceleration of the plasma. The first order quantities read:
\begin{equation}
\frac{\partial \delta \rho}{\partial t} + \rho_0\vec{\nabla}\cdot (\delta \vec{v}) = 0 
\label{eq:cont2}
\end{equation}
\begin{eqnarray}
\rho_0\frac{\partial \delta \vec{v}}{\partial t} &+& c_{s}^2 \nabla \delta \rho  - \frac{\delta \rho}{\rho_0} \frac{\vec{J}_{cr}}{c} \times \vec{B}_0  \nonumber \\
&-&\frac{1}{4\pi}(\vec{\nabla} \times \vec{\delta B})\times\vec{B}_0  + \frac{\vec{J}_{cr}}{c} \times \delta \vec{B} =  0 
\label{eq:prob2}
\end{eqnarray}
\begin{equation}
\frac{\partial \delta \vec{B}}{\partial t} - \nabla \times (\delta \vec{v} \times \vec{B}_0) = 0,
\label{eq:law2}
\end{equation}
with $c_s$ being the sound speed in the plasma and where we have chosen a reference frame where $|\vec{v}_0(t=0)| = 0$. 

\begin{figure}
\includegraphics[width=8.6cm]{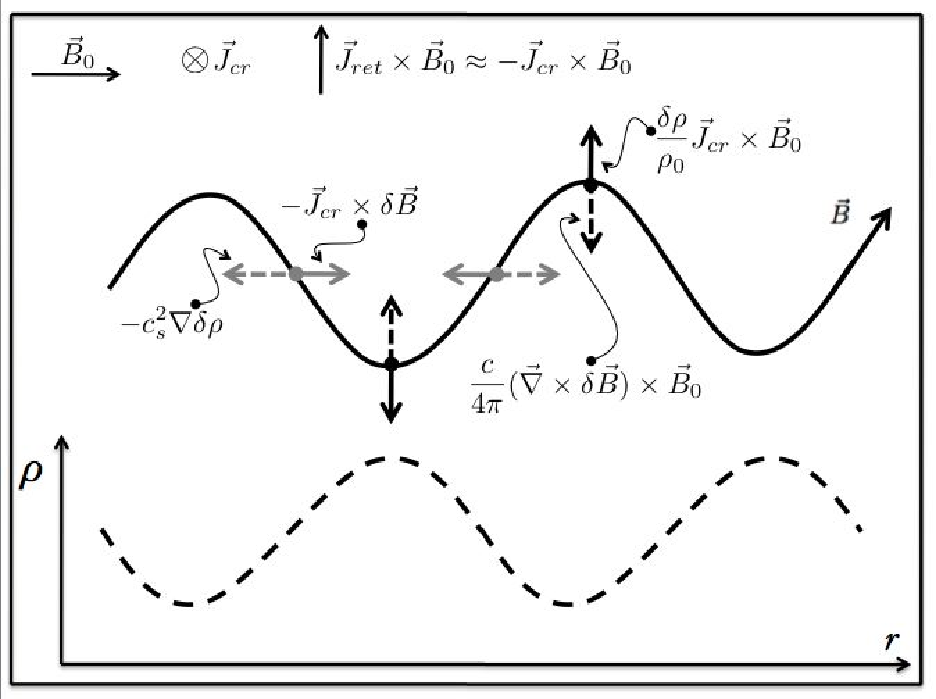}
\caption{A diagram illustrating the physics of PCDI. If an external current $\vec{J}_{cr}$ is applied perpendicular to the plane of the figure and also perpendicular to the initial field, $\vec{B}_0$, the background plasma will develop a return current, $\vec{J}_{ret} \approx -\vec{J}_{cr}$ pointing out of the plane of the figure. This will produce a homogeneous force $-\vec{J}_{cr} \times \vec{B}_0$ accelerating the plasma, as shown in the diagram. The PCDI begins from this non-equilibrium state. Let us consider the magnetic field lines and the plasma density are represented by the solid and dotted lines of the figure, respectively. Then, since the force $-\vec{J}_{cr} \times \vec{B}_0$ is the same everywhere, regions of low density will accelerate more than regions of high density, making the overdense regions lag behind with respect to the rest of the plasma. This is the origin of the force $\delta \rho/\rho_0\vec{J}_{cr} \times \vec{B}_0$ of Eq. (\ref{eq:prob2}) (shown by the solid, black arrows), which tends to stretch the lines of magnetic field and amplify $\delta \vec{B}$. The other forces from the same equation: $-\vec{J}_{cr} \times \delta \vec{B}$, $-c_{s}^2 \nabla \delta \rho$, and $(\vec{\nabla} \times \vec{\delta B})\times\vec{B}_0/4\pi$, are represented by the solid grey arrow,  the dotted grey arrow, and the dotted black arrow, respectively. For unstable modes, this amplification happens because the growth of $\delta \vec{B}$ increases the force $-\vec{J}_{cr} \times \delta \vec{B}$, which enhances the density fluctuations. Thus, the growth of $\delta \vec{B}$ increases the force that produces it, resulting in an exponential growth of $\delta \vec{B}$ and $\delta \rho$.}
 \label{fig:figure1}
 \end{figure}

The physical meaning of the different force terms in Equation (\ref{eq:prob2}) can be understood with the help of Figure \ref{fig:figure1}. Let perturbations $\delta \vec{B}$ and $\delta \rho$ be such that the magnetic field lines and plasma density are represented by the solid and dotted lines of Figure \ref{fig:figure1}, respectively. Then, since the force $\vec{J}_{ret} \times \vec{B}_0$ is the same everywhere, regions of low density will accelerate more than regions of high density, making the overdense regions lag behind with respect to the rest of the plasma. This is the origin of the force $\delta \rho/\rho_0\vec{J}_{cr} \times \vec{B}_0$ in Equation (\ref{eq:prob2}). As shown by the solid black arrows in Figure \ref{fig:figure1}, this force tends to stretch the lines of magnetic field and, therefore, produces the amplification of $\delta \vec{B}$. This amplification can happen exponentially, because the growth of $\delta \vec{B}$ increases the force $-\vec{J}_{cr} \times \delta \vec{B}$, which, as depicted in Figure \ref{fig:figure1}, enhances the density contrast even more. In a sense, the PCDI is similar to Parker instability, since the magnetic growth implies the existence of regions of the plasma that accelerate more than others in the direction perpendicular to $\vec{B}_0$. However, whereas in the case of Parker instability the amplification is driven by a homogeneous gravitational {\it acceleration}, the amplification of PCDI is caused by a homogeneous {\it force}, given by $-\vec{J}_{cr} \times \vec{B}_0$.

Performing the Fourier transform of Eqs. (\ref{eq:cont2}), (\ref{eq:prob2}), and (\ref{eq:law2}), and assuming that the Fourier modes are stationary, with $\vec{k} || \vec{B}_0$, it is straightforward to obtain the dispersion relation:
\begin{equation}
\gamma^2\frac{c^2}{v_A^2} + c^2k^2 = \frac{4\pi J_{cr}^2}{\rho_0(\gamma^2 + c_{s}^2k^2)},
\label{eq:dispersion}
\end{equation} 
where $\gamma$ is the growth rate corresponding to a given $k$. This dispersion relation can also be obtained by taking the limit $\vec{k} || \vec{B}_0 \perp \vec{J}_{cr}$ of the more general MHD result shown in Eq. (4) of \cite{Bell05}. A derivation using the multifluid approach can be found in Appendix (\ref{sec:appendixa}). 
From Equation (\ref{eq:dispersion}) we obtain that the fastest growing mode has the wave number,
\begin{equation}
k_{PCDI}= \frac{4\pi J_{cr}}{c^2B_0} \frac{v_{A,0}^2/c_{s}}{1+v_{A,0}/c_{s}}\frac{c}{\sqrt{v_{A,0}c_{s}}},
\label{eq:kmax}
\end{equation}
and the growth rate,
\begin{equation}
\gamma_{PCDI} = 2J_{cr} (\pi/\rho c^2)^{1/2} \frac{v_{A,0}/c_{s}}{1+v_{A,0}/c_{s}},
\label{eq:gammamax}
\end{equation}
where $v_{A,0}$($\equiv B_0/\sqrt{4\pi\rho_0}$) is the initial Alfv\'{e}n velocity in the plasma. Thus, if the background plasma is initially cold ($c_s \to 0$), then $k_{PCDI} \to \infty$ and $\gamma_{PCDI} \approx 2J_{cr} (\pi/\rho c^2)^{1/2}$. However, if we assume that $c_s$ is close to the typical turbulent velocity in the plasma, then it is possible to show from Eqns. \ref{eq:cont2}-\ref{eq:law2} that the sound speed due to PCDI turbulence would be $\sim v_{A,0}\delta B/B_0$. Thus, the thermal effects would make $k_{PCDI} \approx 2\pi J_{cr}/c\delta B$ and would keep the same growth rate $\gamma_{PCDI} \approx 2J_{cr} (\pi/\rho c^2)^{1/2}$ as long as $c_s \lesssim v_{A,0}$.

As $\delta B/B_0 \to 1$, the expressions for the wave number of fastest growth, $k_{PCDI}$, and its corresponding growth rate, $\gamma_{PCDI}$, look the same as the ones for CRCD instability: $k_{CRCD} = 2\pi J_{cr}/cB_0$ and $\gamma_{CRCD}=J_{cr} (\pi/\rho c^2)^{1/2}$ \citep{Bell04}. However, the CR currents that drive the PCDI and CRCD instabilities are the components of $\vec{J}_{cr}$ perpendicular and parallel to $\vec{B}_{0}$, respectively. Thus, if $\vec{J}_{cr}$ is quasi-perpendicular to $\vec{B}_0$ (as may be the case in regions close to the shock), the PCDI would grow faster than the CRCD instability.

\section{PIC Simulations}
\label{sec:simulations}
In this section we use two-dimensional PIC simulations to study the PCDI. First, in \S \ref{sec:constcurr}, we explore the case 
of a constant, externally imposed CR current, $\vec{J}_{cr}$, perpendicular to the ambient field, $\vec{B}_0$. Then, in \S \ref{sec:crback}, we study the full dynamical evolution of the CRs, allowing a non-constant $\vec{J}_{cr}$. We pay special attention to the way a current $\vec{J}_{cr}$ perpendicular to $\vec{B}_0$ can set in, and to the effect of the CR back-reaction on the saturation of the instability. In each section we specify the numerical set up of the corresponding simulations, listing their common features in \S \ref{sec:common}.
 
 \subsection{Common Simulation Setup}
 \label{sec:common}
We use the electromagnetic PIC code TRISTAN-MP (Buneman 1993; Spitkovsky 2005). Our simulations use periodic, two-dimensional boxes whose normal vector is chosen to point along $\hat{z}$. 
The boxes contain a charge- and current-neutral plasma, consisting of cold ions and electrons (with initial thermal velocities $\sim 10^{-4}c$), which constitute the background plasma, and a population of relativistic particles or CRs that may either be composed of ion-CRs (positive CRs) only, or a mixture of both ion- and electron-CRs. The boxes also contain the initial magnetic field, $\vec{B}_0$, in the plane of the simulation, pointing along the $\hat{y}$ axis. Particles are spread within the box in such a way that the local charge density is zero everywhere. Since typically in a SNR the CR density $n_{cr}$ is much smaller than the background ion density, $n_i$, we increase CR statistics by modifying the charge of the plasma and CR macroparticles (for instance, by making $Q_{i,cr} \sim Q_in_{i,cr}/n_{i}$, where $Q_{i}$ and $Q_{i,cr}$ are the charge of the ion and ion-CR macroparticles). The mass of the macroparticles is modified accordingly in order to keep their charge to mass ratios unchanged.
 
\subsection{Constant CR current}
 \label{sec:constcurr}
In this section we focus on the case of a constant $\vec{J}_{cr}$ pointing along $\hat{z}$ (out of the plane of the simulation). Thus, the back-reaction of the field on the CR trajectories is explicitly suppressed in these simulations. We show the results of four runs with different $n_{cr}/n_i$ ratios and CR charge densities, so that we can test the dispersion relation (Eq. \ref{eq:dispersion}) and investigate the effect of having different CR net charge. In two of our simulations the CRs have no net charge (they are made of equal amount of ion- and electron-CRs), while in the other two runs CRs are only positively charged. In the case of zero CR net charge, the CR current is produced by two counter-streaming CR beams of equal density but opposite charge. For both the zero and non-zero CR charge cases we test $n_{cr}/n_i = 0.01$ and 0.04, where $n_{cr}$ is the total density of CRs (which in the case of zero CR charge includes the positive and negative CRs)\footnote{In the zero CR charge case, the charge and number densities of macroparticles we use are $Q_{i,cr}=-Q_{e,cr}= (Q_i/2)n_{cr}/n_i = -(Q_e/2)n_{cr}/n_i$, and $N_{i,cr}=N_{e,cr}=N_{i}=N_{e}$ (where $N_{j}$ represents the number of macroparticle per cell for species ``$j$"). In the non-zero case, $Q_{cr} = -Q_en_{cr}/n_i$. Then, in order to have local charge neutrality we need $Q_{i} = -Q_e(1-n_{cr}/n_i)$, and $N_{cr}=N_{i}=N_{e}$ (this way there is a small excess of electron charge in the background plasma, which compensates the charge of the CRs).}. In both types of simulations, particles are randomly distributed in the box, but placed such that, initially, there is local charge neutrality. Also, the initial local current is made zero by giving electrons a small initial velocity so that they carry a compensating current $-\vec{J}_{cr}$. The other physical and numerical parameters of the simulations are: $v_{A,0}/c=1/40$, mass ratio $m_i/m_e=10$, CR current velocity $v_{d,cr}=0.6c$, speed of light $c=0.45\Delta/\Delta t$, the skin depth of electrons $c/\omega_{p,e}=3 \Delta$, and the number of macroparticles per cell per species is 16. Here, $\Delta$ is the grid cell size and $\Delta t$ is the time step.

\begin{figure}
\centering
\includegraphics[width=8.5cm]{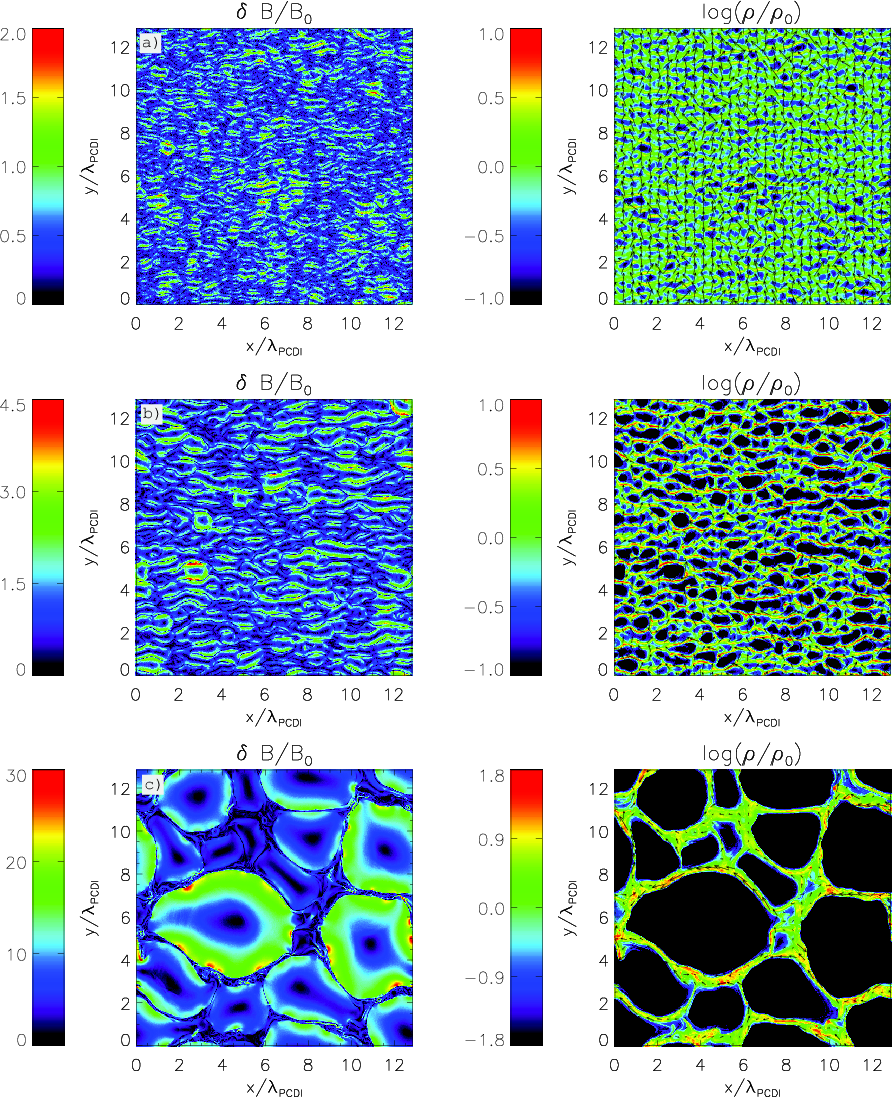}
\caption{The space distribution of the magnitude of the generated field and the background plasma density at $t\gamma_{PCDI} = 17, 21$, and 35 (represented by panels a), b), and c), respectively), for the run with $n_{cr}/n_i=0.01$ and neutral CR beam. The initial magnetic field, $\vec{B}_0$, and the CR current, $\vec{J}_{cr}$, point along $\hat{y}$ and $\hat{z}$, respectively. The distance is normalized in terms of the theoretical $\lambda_{PCDI}$ at $\delta B/B_0 = 1$. We see that in the linear regime (panel a) the size of the fluctuations, $\lambda$, is smaller than $\lambda_{PCDI}$, and that it evolves into $\lambda \sim (\delta B/B_0)\lambda_{PCDI}$ as the amplification becomes non-linear (panels b) and c). When $\delta B/B_0 \gg 1$ (panel c), the PCDI modes acquire a loop-like shape.} 
\label{fig:magden} 
\end{figure}

First, we tested the dominant length scale for the PCDI amplification. Figure \ref{fig:magden} shows the space distribution of the field and background plasma density for the run with $n_{cr}/n_i=0.01$ and neutral CR beam at three different times ($t\gamma_{PCDI} = 17, 21$, and 35, where $\gamma_{PCDI}$ is the theoretical growth rate of the simulation). When $\delta B/B_0 < 1$ ($t\gamma_{PCD} = 17$, in panel $a$), the dominant wavelength is smaller than $\lambda_{PCDI}$ (which is chosen to be the wavelength of the fastest growing mode when $\delta B/B_0 = 1$). When $\delta B/B_0 \approx 1$ ($t\gamma_{PCD} = 21$, in panel $b$), the length scale of the fluctuations becomes very close to $\lambda_{PCDI}$, confirming our analytical result. As the field becomes non-linear, it adopts the shape of loops, which tend to merge and grow in size. We see that in the very non-linear regime ($t\gamma_{PCD} = 35$) the dominant scale of the loops grows by a factor comparable to the magnitude of the amplification, which is consistent with our predictions.

\begin{figure}
\includegraphics[width=8.5cm]{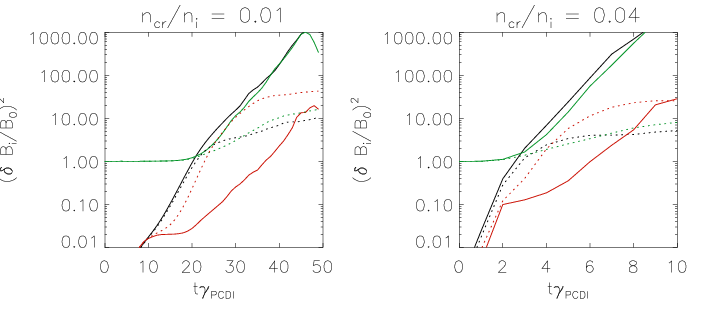}
 \caption{The evolution of the magnetic energy in the three cartesian axes (black, green, and red lines correspond to the $x$, $y$, and $z$ axes, respectively) for the four simulations in \S \ref{sec:constcurr}. The left and right panels show the cases with $n_{cr}/n_{i}=0.01$ and 0.04, respectively. The solid lines show the situations when the background plasma (and the CRs) have a zero net charge, while the dotted lines represent the cases when the background plasma has a negative net charge compensated by the positive CRs. The time is measured in units of $\gamma_{PCDI}^{-1}$, where $\gamma_{PCDI}$ is the theoretical growth rate of the instability for the case $n_{cr}/n_{i}=0.04$.}
 \label{fig:magenergy}
 \end{figure}

Our simulations also confirm the analytic growth rate, given in Eq. (\ref{eq:gammamax}). Figure \ref{fig:magenergy} shows the evolution of the energy in the three magnetic field components in our simulations. The left and right panels represent the cases with $n_{cr}/n_{i}=0.01$ and $0.04$, respectively. Solid and dotted lines show the cases with zero and finite CR charge. The colors black, green, and red represent the magnetic energy of the $x$, $y$, and $z$ field components. The time is measured in units of $\gamma_{PCDI}^{-1}$, where $\gamma_{PCDI}$ is the PCDI theoretical growth rate for $n_{cr}/n_{i}=0.04$. In the linear regime, the PIC results are consistent with our theoretical estimates for the PCDI growth rates. Also, we see that, as long as $\delta B \ll B_0$, the growth rate is independent of the net CR charge. This situation changes in the non-linear regime. While in the zero CR charge cases the magnetic field keeps growing with about the same growth rate as in the linear regime, the non-zero CR charge simulations saturate at $\delta B/B \sim 1$. 

The non-linear behavior of PCDI can be understood as follows.
As the amplified field becomes non-linear, it adopts the shape of magnetic loops that surround background plasma holes on scales of $\sim \lambda_{PCDI}$ (as shown in Figure \ref{fig:magden}). The evolution of magnetic loops in the presence of a constant, perpendicular CR current was already considered by \cite{MilosavljevicEtAl06}, whose argument we revise here. If $\vec{J}_{cr}$ is constant, the background plasma will create a return current, $\vec{J}_{ret} \approx -\vec{J}_{cr}$. Thus, this return current will produce a force $\approx -\vec{J}_{cr} \times \vec{B}$ that tends to expand the loops. In the MHD regime, as a loop expands, the magnetic flux freezes in the fluid, keeping the ratio $B/\rho r$ constant (where $r$ is the radius of the loop). Thus, if no extra forces are considered, we obtain that the acceleration of the loop is $d^2\vec{r}/dt^2 \sim J_{cr}B_0\vec{r}/\lambda_{PCDI}\rho_0c$, which is solved by $r(t) \sim \lambda_{PCDI}e^{\gamma_{PCDI} t}$, where $\lambda_{PCDI}$ is equal to $2\pi/k_{PCDI}$ in the limit $\delta B=B_0$. Thus, since flux freezing keeps $B/\rho r$ constant, the magnetic field should also grow exponentially at approximately the same rate as $r(t)$.
However, we see that, when CRs are made of only positively charged particles, this non-linear amplification gets suppressed. In that case, the expansion of the negatively charged loops would produce charge separation in the plasma. Due to this, a monopolar electric field will form in the loops, which would suppress their expansion.  Although this force plays an important role in the cases analyzed in this section, in cases where $\vec{J}_{cr}$ is caused by CRs protruding into regions of pre-amplified field (like in the lower plot of Figure \ref{fig:figure2}), they would not stop the growth. The reason has to do with the particular way the charges are distributed within each loop. We will analyze in detail this point in Appendix \ref{app:B}.
 
In summary, in this section we have used two-dimensional PIC simulations to show that, if the CR current is kept constant, the PCDI grows at the wavelengths and rates predicted by the linear theory. In the non-linear regime, the exponential growth can continue with about the same rate as in the linear regime, and with a length scale proportional to $\delta B$. However, if CRs are mainly positively charged particles, this growth may be affected by electrostatic fields due to charge separation in the plasma (see Appendix \ref{app:B}).  
     
\subsection{CR back-reaction}
\label{sec:crback}
In this section, we include the full dynamics of CRs our the study. In \S \ref{sec:protru}, we show how CRs protruding into regions of pre-amplified field can produce a $\vec{J}_{cr} \perp \vec{B}_0$, and drive the growth of PCDI modes. Then, in \S \ref{sec:saturation} we study the effect of the CR back-reaction on the saturation of the instability.

\subsubsection{Growth of PCDI}
\label{sec:protru}
In our simulations we want to reproduce the situation pictured in Figure \ref{fig:figure2}, where regions with a pre-amplified field (represented  by grey ``doughnut-like" shapes) are surrounded by a population of magnetized CRs (i.e., CRs with Larmor radii smaller than the length scales of these regions). In particular, we want to model small sections of these regions, like the one marked with a dotted-line box in the upper panel of Figure \ref{fig:figure2}. The idea is to study the growth of the PCDI due to CRs protruding into the regions of pre-amplified field region. Since we use a monoenergetic population of CRs, these particles are expected to penetrate by a distance comparable to two times their Larmor radius, $R_{L,cr}$, which defines a region or interface where the field grows. The other parts of the plasma, which are not reached by these CRs, should experience compression due to the acceleration of the interface. As CRs turn in the field of the magnetized region, the acquire the mean velocity in the $z$ direction (out of the box), and the associated perpendicular current drives the rippling of the field characteristic of the PCDI. 

The numerical setup we use is similar to the one of the constant CR current case described in \S \ref{sec:common} and \ref{sec:constcurr}. The only difference is that, in this case, CRs are positively charged and monoenergetic particles, and are isotropically injected within a narrow strip along the $\hat{y}$ axis. This strip is located in the left part of the simulation box and its width is $1/40$th of the width of the box. Thus, whereas the strip represents the region with high CR density and low initial magnetic field, the rest of the simulation box would represent the inner part of the region of pre-amplified field. Extra electrons are injected along with the CRs so that the net injected charge is zero. The injection criterion is such that the mean density of CRs in the strip, $n_{cr,0}$, is kept constant\footnote{Similarly to the case with constant CR current, the charge of the macroparticles is modified such that it is possible to have a large number of CR macroparticles per cell. So, $Q_{cr} = Q_in_{cr,0}/n_{i,0}$, where $n_{i,0}$ is the physical density of ions in the strip at $t=0$.}. 
\begin{figure}
\includegraphics[width=8cm]{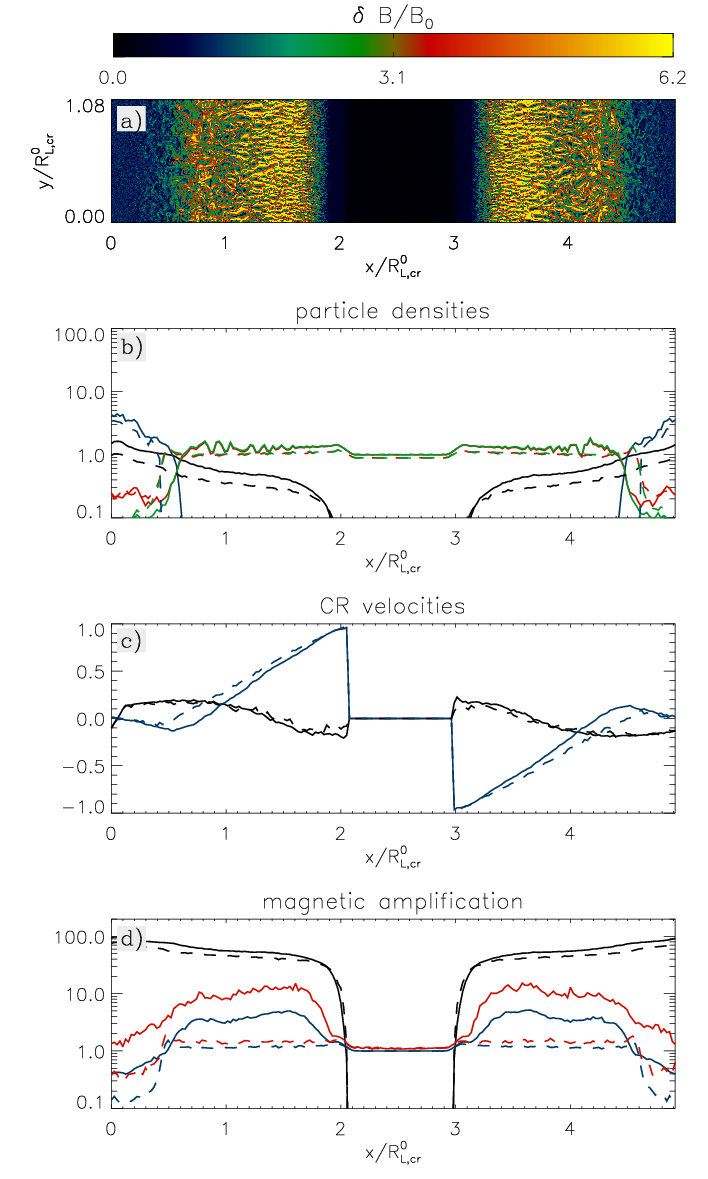}
\caption{Distribution of different particle and field quantities for simulation $S_*$ at $t\gamma_{PCDI} = 32$, where $\gamma_{PCDI}$ is the growth rate of the PCDI calculated using $n_{cr,0}$ and $c/2$ as the number density and mean velocity of CRs. The distance is shown in units of $R_{L,cr}^0$, which is the Larmor radius of the CRs in the initial field $\vec{B}_0$ ($=B_0\hat{y}$). The simulation box is periodic. a) Magnitude of the generated field; b) Density of different species averaged over $y$ as a function of the $x$ coordinate. The red and green lines represent the background ions and electrons, whose densities are normalized by $n_{i,0}$, and the black and blue lines represent the densities of CRs and injected electrons (to compensate CR charge), which are normalized in terms of $n_{cr,0}$; c) Mean velocities of CRs along $\hat{x}$ (black line) and $\hat{z}$ (blue line), normalized by $c$; d) Magnetic field amplification as a function of $x$, normalized in terms of $B_0$. The red and blue lines show the maximum and average amplifications, respectively. Black line shows the field magnitude corresponding to local energy equipartition with the CRs. Dashed lines in panels b), c) and d) show the results of a simulation analogous to $S_*$, but with the $y$ dimension of the box significantly reduced (quasi-1D). This simulation shows only the effect of compressional amplification.}
\label{fig:protru32}
\end{figure}
\begin{figure}
\includegraphics[width=8cm]{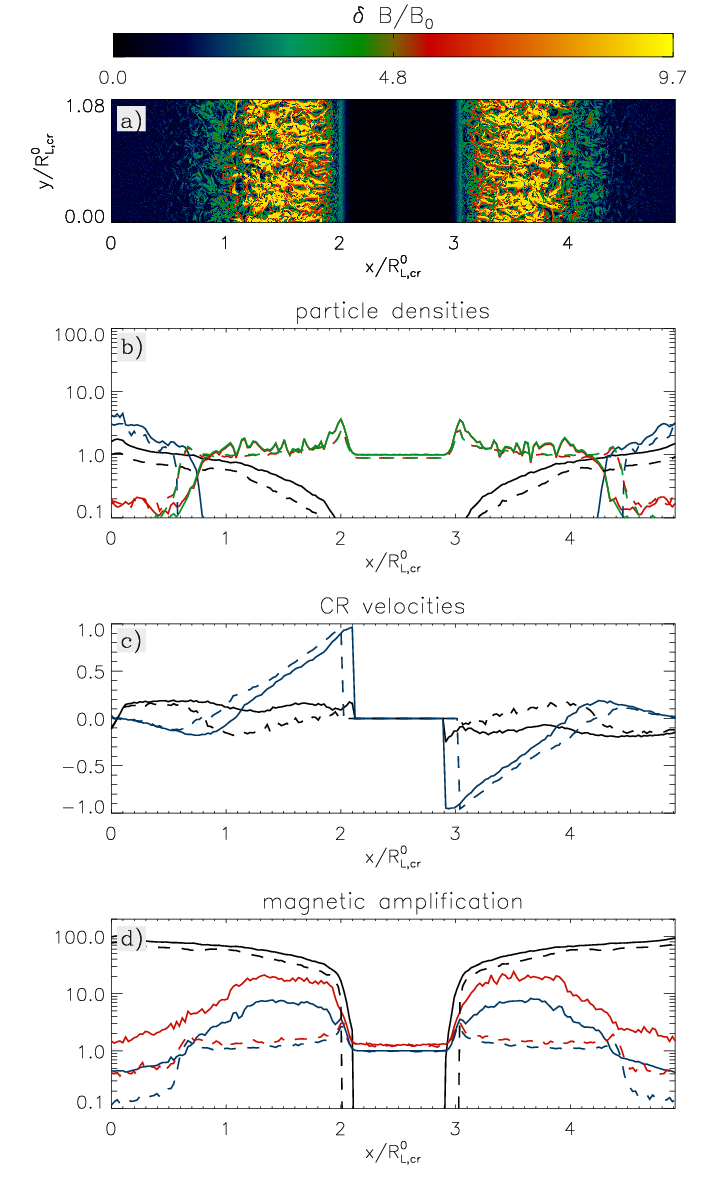}
\caption{Same as Figure \ref{fig:protru32}, but at $t\gamma_{PCDI} = 40$.}
\label{fig:protru40}
\end{figure} 
\begin{figure}
\includegraphics[width=8cm]{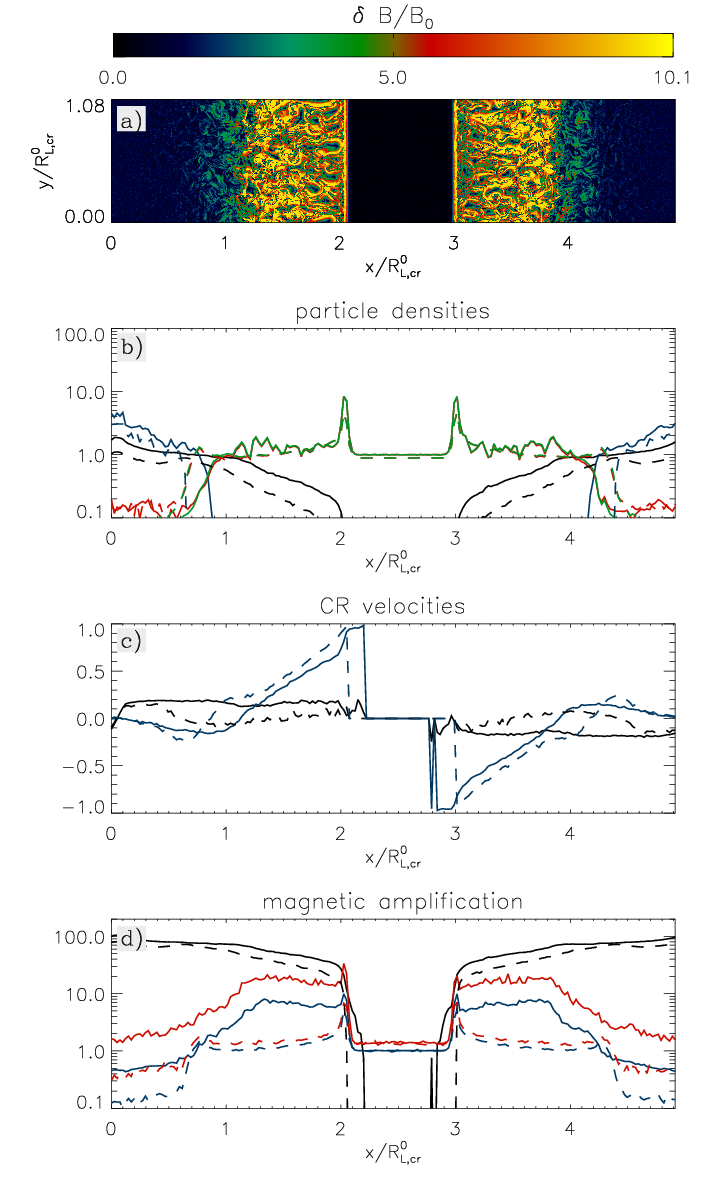}
\caption{Same as Figure \ref{fig:protru32}, but at $t\gamma_{PCDI} = 44$.}
\label{fig:protru44}
\end{figure}  

We ran a series of simulations that share the following parameters: $m_i/m_e=10$, $n_{cr,0}/n_{i,0}=0.04$ (where $n_{i,0}$ is the initial background ion density in the injection strip), $\Gamma_{cr}=5$, $c=0.45\Delta/\Delta t$, $c/\omega_{p,e} = 6.4$, and use 8 macroparticle per cell per species. The difference between the runs is given by their initial
Alfv\'{e}n velocity, $v_{A,0}/c$, which ranges from $v_{A,0}/c=1/160$ to 1/40. The evolution of the relevant physical quantities for a simulation with $v_{A,0}/c=1/80$, which we will call ``$S_*$", is depicted in Figures \ref{fig:protru32}, \ref{fig:protru40}, and \ref{fig:protru44}. These figures correspond to $t\gamma_{PCDI} = 32, 40$ and 44, where $\gamma_{PCDI}$ is the growth rate of the PCDI considering $n_{cr} = n_{cr,0}$ and $v_{d,cr} = c/2$.      

In panels $a)$ of these Figures, we depict the space structure of the generated field. As CRs penetrate into the regions of zero initial CR density, we see the formation of magnetic loops, similar to the ones shown in the case of constant $\vec{J}_{cr}$ (\S \ref{sec:constcurr})\footnote{Since the simulation box is periodic, CR propagate both to the left and to the right of the injection strip. Thus, two nearly symmetric regions of field amplification form on both ends of the box.}. In panels $b)$ we show the density of the different species as a function of the $x$ coordinate. It shows how CRs can penetrate into the magnetized region by a distance $\sim 2R_{L,cr}$, while the injected electrons (necessary to compensate the CR charge) stay close to the injection strip\footnote{The fact that CRs can only penetrate by a single distance given by $\sim 2R_{L,cr}$ is caused by our particular choice of a monoenergetic CR energy distribution. In reality, CRs should have a wide distribution of energies, so they should be able to penetrate by different distances, depending on their energy. We will get back to this point at the end of \S \ref{sec:crback}.}. The plots $c)$ represent the CR mean velocities along $\hat{x}$ and $\hat{z}$ (pointing out of the simulation plane). We can see how the mean CR velocity is dominated by its $z$ component (in the three Figures of the order of $\sim c/2$), due to their deflection in the $B_y$ field, which confirms the model presented in the lower plot of Figure \ref{fig:figure2}. Finally, the plots $d)$ show the mean and maximum magnetic amplification as a function of $x$.
We can see that the typical wavelengths of the loops are consistent with the analytical theory, which predicts the dominant wavelength to be such that $\lambda_{PCDI}/R^{0}_{L,cr} \approx (v_{A,0}/c)^{2}(8\pi n_i/n_{cr}\Gamma_{cr})(\delta B/B_0)$, where we have assumed that the mean CR velocity along $z$ is $\sim c/2$. Applying this formula to the cases of Figures \ref{fig:protru32}, \ref{fig:protru40}, and \ref{fig:protru44}, we see that the dominant length scales of the fluctuations are about $1.5$ times smaller than what the theory predicts.
Also, as anticipated at the end of \S \ref{sec:constcurr}, the charge separation due to loop expansion does not stop the non-linear growth of the PCDI in this case. The reason is related to the way the background plasma compensates the charge of CRs, as they penetrate into regions of low initial CR density. We refer the reader to Appendix \ref{app:B} for details on this point.
\begin{figure}
\includegraphics[width=9cm]{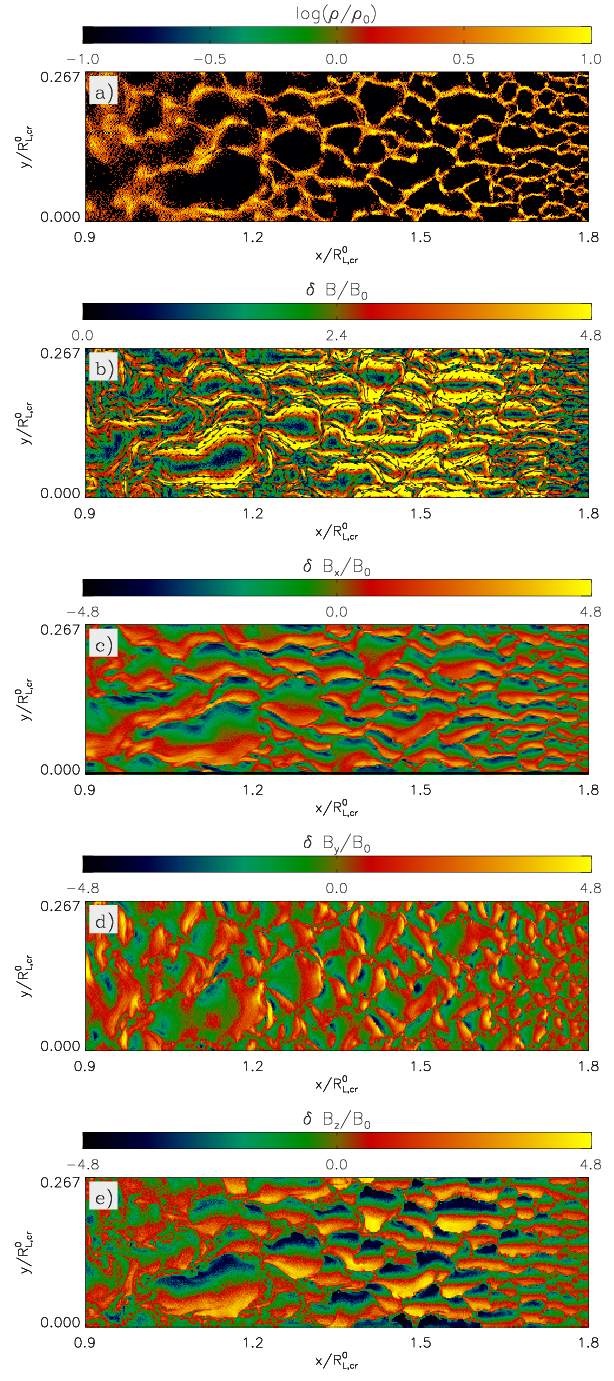}
\caption{The spatial structure of the density fluctuations ($a$), magnetic field strength ($b$), and the $x$, $y$, and $z$ components of $\vec{B}$ (plots $c$, $d$, and $e$) are depicted for a region of simulation $S_*$ at $t\gamma_{PCDI}=32$. The arrows in plot $b)$ represent the field projectin on the plane $z=0$. The density and the field are normalized in terms of $\rho_0$ and $B_0$, respectively.}
\label{fig:protruden}
\end{figure} 

A clearer view of the spatial structure of the PCDI can be seen from Figure \ref{fig:protruden}, which shows the components of the field and the density fluctuations for a selected region of simulation $S_*$ at $t\gamma_{PCDI}=32$. The observed features of the PCDI turbulence are essentially the same as for the case of the externally imposed current: the non-linear field adopts a loop-like structure with very strong density contrasts ($\delta \rho/\rho_0 \sim 10$). 

As was discussed before, the PCDI happens while the background plasma is being pushed by the $-\vec{J}_{cr} \times\vec{B}_0$ force. This pushing can also amplify the field due to a simple field compression. In order to differentiate the growth due to the compression from the one due to the PCDI modes, the dashed lines in Figs. \ref{fig:protru32}, \ref{fig:protru40}, and \ref{fig:protru44} show the results of an analogous simulation where the $y$ dimension of the box has been significantly reduced. This simulation behaves as a one-dimensional run, where the PCDI modes (whose $\vec{k}||\hat{y}$) cannot grow. We see that the magnetic amplification corresponding to this one-dimensional case is significantly smaller than the one due to the growth of the PCDI.

\subsubsection{PCDI saturation}
\label{sec:saturation}
Our simulations show that, similarly to the case of the CRCD instability, the field saturation happens when the Larmor radii of the CRs, $R_{L,cr}$, become comparable to the size of the generated magnetic fluctuations. In our fiducial simulation $S_*$, the magnetic field reaches saturation at $t\gamma_{PCDI} = 40$ (Figure \ref{fig:protru40}), where it is amplified by a factor of $\sim 10$. This implies that the Larmor radii of the CRs,$R_{L,cr}$, should be $\sim 1/10$th of its initial value, $R^0_{L,cr}$. This coincides with the typical length scale of the magnetic fluctuations observed in Figure \ref{fig:protru40}, where the distance is measured in units of $R^0_{L,cr}$. This saturation criterion can be understood in the same way as for the case of the CRCD instability. Since the magnetic growth is driven by the CR current (in this case perpendicular to $\vec{B_0}$), the growth will stop when this current is significantly reduced, which happens when CR trajectories are strongly perturbed by the generated field. This requires the length scale of the fluctuations to be comparable to $R_{L,cr}$. We check this in Figure \ref{fig:protru44}, where the $z$ component of the CR mean velocity (in regions where $n_{cr} > 0.1n_{cr,0}$) has been reduced to almost half of its value in Figure \ref{fig:protru40}. 

By considering the expected evolution of the size of the magnetic fluctuations ($\approx 2\pi/k_{PCDI} \approx cB/J_{cr}$) and assuming that the CR mean velocity perpendicular to the field is $\sim c/2$, it is possible to estimate the maximum field reachable by the PCDI. If we make $cB/J_{cr} \approx R_{L,cr} = m_{cr}c^2\Gamma_{cr}/e_{cr}B$, where $m_{cr}$ and $e_{cr}$ are the mass and charge of the CRs, its possible to show that saturation will happen at $B^2/8\pi \approx n_{cr}m_{cr}c^2\Gamma_{cr}/16\pi$. This implies that the field at saturation should be a factor of $\sim \sqrt{16\pi} \approx 7$ below energy equipartition with the CRs. We can test this criterion by looking at panels d) of Figs. \ref{fig:protru40} and \ref{fig:protru44}. We see that the maximum amplification is about $\sim 3$ times smaller than the one corresponding to local energy equipartition between CRs and magnetic field.  The mean magnetic amplification is an extra factor of $\sim 3$ smaller, due to the existence of regions of very low density and magnetic field amplitude in the PCDI turbulence. We can also check this condition by seeing the saturation values of the field for simulations with the same CR energy density as in simulation $S_*$, but with different values of $B_0$. 
\begin{figure}
\includegraphics[width=8cm]{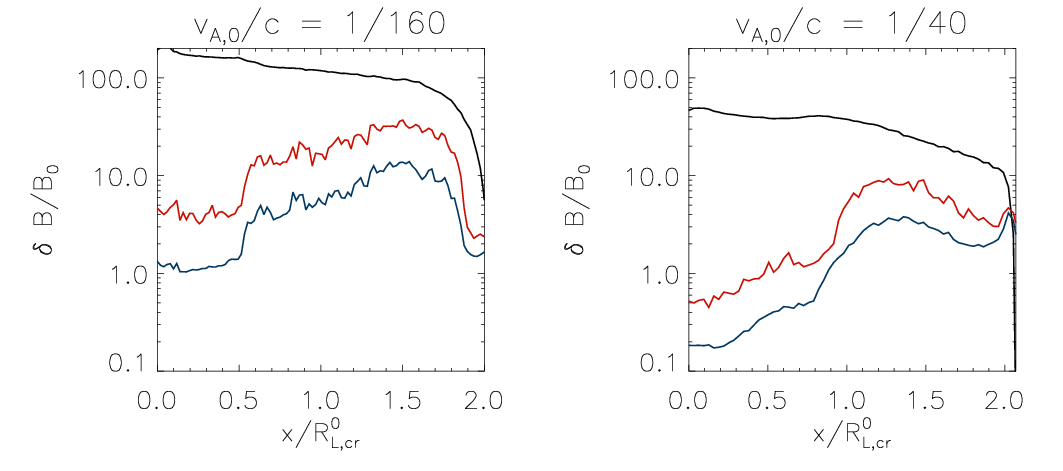}
\caption{The left and right panels show magnetic amplification (normalized in terms of $B_0$) as a function of $x$ at the moment of saturation for simulations analogue to $S_*$, but with $v_{A,0}/c =1/160$ and 1/40, respectively. The red and blue lines show the maximum and average amplification, respectively. And the black line shows the magnetic amplitude of a field at local energy equipartition with the CRs.}
\label{fig:prosatura}
\end{figure} 
The left and right panels in Figure \ref{fig:prosatura} show the mean and maximum magnetic amplification as a function of $x$ at saturation for simulations with $v_{A,0}/c =1/160$ and 1/40, respectively. Again, the observed saturation is consistent with the mean magnetic field being $\sim 7$ times smaller than the one corresponding to energy equipartition with the CRs.

The main uncertainty in this saturation criterion is given by the particular CR energy distribution we use, which corresponds to monoenergetic CRs. This implies that CRs are able to penetrate into the simulation box by a distance shorter than $\sim 2R_{L,cr}$. This defines a region or ``interface" where PCDI modes grow, with the rest of the plasma only being compressed due to the acceleration of the plasma in the interface. The beginning of the compression of the central part of the region of pre-amplified field can be seen in Figure \ref{fig:protru44}, where the density of the background plasma shows prominent spikes at $x/R_{L,cr}^0\approx 2$ and 3. However, in reality, CRs should have a wide distribution of energies. So, in principle, the magnetized CRs (the ones with $R_{L,cr}\lesssim \lambda_{0}$) should be able to fill most of the volume of pre-amplified field. Also, since the number density of the CRs is expected to be a strongly decreasing function of their energy, low energy CRs (with small $R_{L,cr}$ and, therefore, filling a small fraction of the volume of the pre-amplified field region) will produce PCDI modes on length and time scales significantly shorter than the ones of the highest energy CRs (that still satisfy the magnetization condition, $R_{L,cr}\lesssim \lambda_{0}$).  Thus, in a realistic setup, the growth of PCDI field should start at the boundary of the regions of pre-amplified field and then propagate towards their interior due to the action of higher energy particles. By choosing monoenergetic CRs, we ignore this effect, and concentrate on capturing the basic physics of field amplification by the PCDI instability, including the growth of PCDI modes due to the formation of a $\vec{J}_{cr}$ perpendicular to $\vec{B}_{0}$, and the effect of the CR back-reaction on the final saturation of the instability.  A more detailed study including the effect of a broad CR energy distribution will be presented elsewhere.

\section{Application to SNR shocks}
\label{sec:discussion}
We have found that the saturation of the PCDI is determined by the CR back-reaction, and that it would imply magnetic fields $\sim \sqrt{16\pi} \approx 7$ times below energy equipartition with the CRs. We can now apply this result to the expected conditions in the upstream medium of SNR shocks. If we assume that all CR energy  is carried by a monoenergetic CR population, we can estimate the maximum amplification factor given by
\begin{equation}
(\delta B/B_0)_{max} \approx 45 \frac{v_{sh}}{10^4 \textrm{km/s}}\frac{10 \textrm{km/s}}{v_{A,0}}\Big(\frac{\eta}{0.05}\Big)^{1/2},
\label{eq:estimate}
\end{equation}
where $\eta \equiv u_{cr}/u_{th}$ is the ratio between the energies of the CRs, $u_{cr}$, and the one of the downstream thermal plasma, $u_{th}$. The main uncertainty in this estimate is given by the assumption that CRs are monoenergetic, so they will only amplify the field in an interface of width $\sim 2 R_{L,cr}$ within the regions of pre-amplified magnetic field (see lower panel in Figure \ref{fig:figure2}). In reality, CRs will have a wide distribution of energies, so they are expected to enter most of the volume where the field has been previously amplified. Since the density of CRs decreases with their energy, the growth of PCDI modes should happen faster in the outer part of the magnetized regions, and gradually propagate towards the inner parts due to the action of the higher energy CRs. The exact details of how this process unfolds, and how it would change the estimated amplification of Eq. (\ref{eq:estimate}), requires further study, taking into account more realistic CR energy distribution. 

The maximum amplification \ref{eq:estimate} is about 4 times larger than for the case of the CRCD instability. Since the saturation criteria of the two mechanisms are analogous, then the factor of $\sim 4$ is explained by the different mean CR velocities used for calculating the corresponding $\vec{J}_{cr}$ (perpendicular and parallel to $\vec{B}_0$ for the PCDI and CRCD instability, respectively). While for the CRCD instability the mean velocity of the CRs can be considered close to $v_{sh}$, in the case of the PCDI it is $\sim c/2$. Thus, the maximum amplification should differ by a factor $\sqrt{(c/2)/v_{sh}}$ ($\approx 4$ for $v_{sh} = 10^4$km/s). If the estimate given in Eq. (\ref{eq:estimate}) were calculated considering the CRs that are already diffusing with a speed of about $v_{sh}$, then both instabilities would saturate at about the same level.

It is important to make sure that the PCDI would have time to grow before being advected into the shock. The growth of the PCDI field is expected to happen at a distance from the shock that is {\it at least} the distance that CRs protrude into the pre-amplified field, $\sim 2R_{L,cr}$ (this would correspond to the field loop that is the closest to the shock, in Figure \ref{fig:figure2}). Then, we can define an advection time, $t_{adv}$, that would be at least $2R_{L,cr}/v_{sh}$. In addition to that, the typical time scale of growth is given by the inverse of the growth rate of the instability, $\gamma_{PCDI} = (J_{cr}(\pi/\rho c^2))^{1/2}$. Therefore, the product $t_{adv}\gamma_{PCDI} \approx 13 (\eta/0.05) (v_{sh}/10^4 \textrm{km/s})/(4v_{A,0}/10 \textrm{km/s})$ (assuming CRs moving on average at $\sim c/2$) shows that, as long as the growth of the field starts at a distance a few times larger than the typical CR Larmor radius, the advection time would not be a restriction to the growth of the PCDI. 

The PCDI would also contribute to the efficiency of the acceleration of particles at the shock. Indeed, the fact that the field saturation happens at $R_{L,cr} \approx \lambda_{PCDI}$ means that the lower energy CRs would be able to generate magnetic fluctuations at length scales comparable to their Larmor radii, which would increase their diffusion and, therefore, the efficiency of their shock acceleration.

\section{Summary and Conclusions}
\label{sec:conclusions}
In this work we proposed a new plasma instability, the perpendicular current-driven instability (PCDI), as a candidate to amplify magnetic fields due to CRs diffusing in front of SNR shocks. The PCDI consists of purely growing, compressional waves produced by the CR current, $\vec{J}_{cr}$, perpendicular to the initial magnetic field, $\vec{B}_0$. The time and length scale of growth of the PCDI are similar to the ones of the CRCD instability, i.e., $\gamma_{PCDI} \approx J_{cr}(\pi/\rho c^2)^{1/2}$ and $\lambda_{PCDI} \approx cB/J_{cr}$. However, whereas the CRCD instability is driven by a current {\it parallel} to $\vec{B}_0$, the PDCI is produced by a current {\it perpendicular} to the initial field. Thus, the fastest growing instability will be determined by whether $\vec{J}_{cr}$ is quasi-perpendicular or quasi-parallel to the $\vec{B}_0$.

We show that, in the upstream medium of SNR shocks, the required perpendicular current can be due to CRs with Larmor radii, $R_{L,cr}$, smaller than the length scale of a previously amplified magnetic turbulence, $\lambda_0$. This scenario is motivated by earlier PIC studies showing that, far from the shock, the CRCD instability can produce non-linear magnetic amplification, driven by the highest energy CRs. This amplification, although limited to a factor of less than $\sim 10$ for the expected energy carried by CRs, would be characterized by non-linear fluctuations in the plasma density and magnetic field strength \citep{RiquelmeEtAl09}. The typical length scale of these fluctuations would be close to the Larmor radius of the highest energy CRs (which is the saturation criterion that implies a maximum amplification factor of $\sim 10$), producing a situation where most CRs would have a $R_{L,cr}$ significantly smaller than the typical size of the pre-amplified fluctuations. 
We show that, under these conditions, the current perpendicular to $\vec{B}_0$ can be produced by lower energy CRs protruding into the regions of pre-amplified field by a distance $\sim 2R_{L,cr}$, which would give rise to a mean velocity of magnitude $\sim c/2$ perpendicular to the field. Due to this, in regions close to the shock, $\vec{J}_{cr}$ would be quasi-perpendicular to the field, making the PCDI grow faster and on length scales smaller than the scales of the CRCD instability.

We used the two-dimensional PIC simulations to study the evolution of the PCDI. We showed that the instability exists and confirmed its theoretical growth rate. We also studied its possible saturation mechanisms, and found that the strong deflection of CRs in the amplified field ultimately stops the magnetic growth, which happens when the the Larmor radius of the CRs, $R_{L,cr}$, is close to the size of the magnetic fluctuations, $\lambda_{PCDI}$. This saturation mechanism is qualitatively the same as the one found for the CRCD instability \citep{RiquelmeEtAl09}. However, since the perpendicular CR current is due to a CR mean velocity of magnitude $\sim c/2$, then the maximum magnetic field amplification would be $\sim \sqrt{16\pi} \approx 7$ times below energy equipartition with the CRs. This would imply a maximum amplification factor in the upstream medium of the SNR shocks of $\sim 45$. The maximum downstream amplification could increase to $\sim 200$, including the  compression at the shock.

The PCDI would also contribute to the efficiency of the acceleration of low energy particles at the shock, by providing magnetic fluctuation on scales comparable to their Larmor radii, which allows them to diffuse.
Given the geometry of the PCDI (see Figure \ref{fig:figure1}), where the fluctuations in the fields and particle properties happen in a plane perpendicular to $\vec{J}_{cr}$, we believe that our two-dimensional simulations (with $\vec{J}_{cr}$ pointing out of the plane of the simulation) are already capturing the essence of the instability. However, future three-dimensional simulations will be important to investigate possible extra effects that may appear by including the additional dimension.

One interesting signature of the PCDI is that it predicts the amplified field, $\vec{B}$, that satisfies $B^2 \propto v_{sh}^2$, assuming a weak dependence of $\eta$ ($\equiv u_{cr}/u_{th}$) on $v_{sh}$. This relationship contrasts with the one obtained for the CRCD instability acting alone, $B^2 \propto v_{sh}^3$ \citep{Bell04}, and constitutes a possible observational test to shed light on the nature of the magnetic amplification in SNR shocks \citep[see observational results of][]{VolkEtAl05}.

Although we have focused on the case of non-relativistic shocks in SNRs, the PCDI may also be relevant for the case of upstream magnetic amplification in the relativistic shocks of jets and Gamma Ray Bursts. In that case, as seen from the reference frame of the upstream medium, the CRs will only experience a small deflection before being advected into the downstream medium of the shock. Thus, CRs will provide an electric current parallel to the shock normal that may also amplify PCDI modes. The amplification of magnetic field due to the expansion of magnetic loops (as in the non-linear regime of the PCDI) in the upstream medium of GRB shocks was already discussed by \cite{MilosavljevicEtAl06}. They suggest that the growth of the field would saturate when neighboring loops collide. Our simulations show that these collisions do not stop the growth. Instead, they make neighboring loops merge and grow in size. We also found that under certain conditions, electrostatic forces due to the charge separation involved in the expansion of the loops may contribute to the quenching of the nonlinear growth. This possibility requires a dedicated study, which will be presented elsewhere.

In conclusion, we have shown that the PCDI constitutes a viable amplification mechanism that, in combination with the CRCD instability, strengthens the idea of CRs being responsible for the significant fraction of magnetic amplification inferred from SNR shock observations.

\appendix

\section{Appendix A:  Three-fluid Derivation of PCDI Dispersion Relation}
\label{sec:appendixa}
We calculate the dispersion relation corresponding to the growth of magnetic field fluctuations, $\delta\vec{B}$, due to CRs propagating at velocity $\vec{v}_{cr}$ perpendicular to the initial magnetic field, $\vec{B}_0$. We focus on the case $\delta\vec{B} \perp \vec{B}_0$ and $\vec{k} \parallel \vec{B}_0$, where $\vec{k}$ is the wave vector of the Fourier modes of the fluctuations. Ions, electrons, and CR are modeled as three separate fluids with densities, $n_i$, $n_e$, and $n_{cr}$, such that $n_i + n_{cr} = n_e$. Electrons are approximated as massless particles, whereas ions and CRs have finite masses and sound speeds $c_{s,i}$ and $c_{s,cr}$, respectively.

The essential part of the calculation consists of determining the perturbations to the plasma currents, $\delta \vec{J}$, expected in the presence of the  amplified field. To first order, the current perturbation contributed by species ``$j$" is given by $\delta \vec{J}_j = \delta n_j \vec{v}_j + n_j \delta \vec{v}_j$, where $\delta n_j$ and $\delta \vec{v}_j$ represent the perturbations to the density and velocity of species ``$j$". 
In the case of the CRs, these quantities can be calculated from the perturbed part of  the Euler equation,
\begin{equation}
\begin{array}{lll}
m_i \Gamma_{cr} n_{cr} \frac{\partial \delta \vec{v}_{cr}}{\partial t} & = & -c_{s,cr}m_i\Gamma_{cr} \vec{\nabla}\delta n_{cr} + e n_{cr} \delta \vec{E} \\ 
&&+ e n_{cr} \big( \frac{\delta \vec{v}_{cr}}{c} \times \vec{B}_0 + \frac{ \vec{v}_{cr}}{c} \times \delta \vec{B} \big),
\end{array}
\label{eq:euler}
\end{equation}
where $m_i$ is the mass of the ions (assumed equal to the mass of the CRs), $\Gamma_{cr}$ is the Lorentz factor of the CRs, $e$ is the charge of the ions (also, equal to the one of the CRs), and $\delta \vec{E}$ corresponds the the electric field fluctuations. For the kind of fluctuations we are interested in, the relevant currents will be the ones perpendicular to $\vec{B}_0$. If we assume that $\delta n_j$, $\delta \vec{v}_j$, $\delta \vec{E}$, and $\delta \vec{B}$ are stationary waves that grow exponentially at a rate $\gamma$, then
\begin{equation}
\delta \vec{v}_{cr, \perp} = \frac{e}{m_i \Gamma_{cr}}\frac{\big( \gamma\delta\vec{E} + \omega_{c,cr}\delta\vec{E}\times\hat{y}  \big)}{\gamma^2 + \omega_{c,cr}^2},
\end{equation}
where the subscript $\perp$ stands for ``perpendicular", $\omega_{c,cr}$ is the cyclotron frequency of the CRs, and $\hat{y}$ is the axis parallel to $\vec{B}_0$.

The CR density can be obtained from the $y$ component of Equation (\ref{eq:euler}). If we neglect the term proprotional to $\delta \vec{E}$, we obtain
\begin{equation}
\delta n_{cr} = \frac{-en_{cr} \partial_y(\frac{\vec{v}_{cr}}{c}\times \delta \vec{B})_y}{m_i\Gamma_{cr}(\gamma^2 + k^2c_{s,cr}^2)}.
\end{equation}
This approximation is justified if we look at the $y$ component of the Euler equation for electrons,
\begin{equation}
m_en_e\frac{\partial\delta v_{e,y}}{\partial t} = -en_e(\delta E_y + \frac{\vec{v}_e}{c}\times \delta \vec{B}).
\label{eq:eulerelectron}
\end{equation}
Since they are assumed to have zero mass, then $\delta E_y = -\frac{\vec{v}_e}{c}\times \delta \vec{B}$. The unperturbed electron velocity $\vec{v}_e$ is given mainly by the $\vec{E} \times \vec{B}c/B^2$ drift of the background plasma, which we make initially zero by choosing the appropriate frame of reference. 

In the case of ions and electrons, the velocity fluctuations are calculated from the fluctuations of their drift velocities. These drifts are the $\vec{E} \times \vec{B}c/B^2$ drift, which we call $\vec{v}^{(0)}$, and an extra drift due to the time variation of the fields, $\vec{v}^{(1)} = -(m_jc/e_jB^2)(d\vec{v}_0/dt)\times \vec{B}$ \citep[see Appendix A of][]{RiquelmeEtAl09}. Thus, it is possible to show that
\begin{equation}
\delta \vec{v}^{(0)}_{\perp} = \frac{\delta\vec{E}\times\vec{B}}{B^2}c,
\label{eq:drift0}
\end{equation}
and
\begin{equation}
\delta \vec{v}^{(1)}_{\perp} = \frac{m_jc^2}{e_jB^2}\frac{d\delta\vec{E}}{dt},
\label{eq:drift1}
\end{equation}
where $\delta \vec{v}^{(0)}_{\perp}$ affects equally the ions and the electrons, while $\delta \vec{v}^{(1)}_{\perp}$ only affects the ions (given that electrons are massless).

The density fluctuations of the ions can be obtained directly from the $y$ component of their Euler equation:
\begin{equation}
\delta n_i = \frac{-en_i}{m_i(\gamma^2 + k^2c_{s,i}^2)}\frac{\partial(\frac{\vec{v}^{(1)}}{c}\times \delta \vec{B})_y}{\partial y}
\end{equation} 
It can be shown that in the regime in which $|\vec{v}^{(0)}| \ll \vec{v}_{cr}$, the time variation of the zeroth order component of the electric field is given by $d\vec{E}/dt = -4\pi\vec{u}_{cr}n_{cr}ev_{A,0}^2/c^2$, where $v_{A,0}$ is the initial Alfv\'{e}n velocity of the background plasma. Thus, the homogeneous part of $\vec{v}^{(1)}$ can be expressed in terms of the time variations of the electric field to obtain,
\begin{equation}
\delta n_i = \frac{en_{cr}}{m_i(\gamma^2 + k^2c_{s,i}^2)}\frac{\partial(\frac{\vec{v}_{cr}}{c}\times \delta \vec{B})_y}{\partial y}
\label{eq:ionden}
\end{equation}
Since we are choosing a frame where $\delta E_y$ is initially zero, the term $\partial_y \delta E_y$ can also be neglected. Thus, from the Poisson equation we obtain that $\delta n_e \approx \delta n_i + \delta n_{cr}$. This way we have all the density and velocity fluctuations needed to calculate $\delta \vec{J}$. If we are interested in the case where the CR back-reaction can be neglected, the value of $\delta \vec{J}$ can be calculated by taking the limit $\Gamma_{cr} \to \infty $. In that case we obtain that
\begin{equation}
\delta \vec{J} = \frac{1}{4\pi} \frac{c^2}{v_A^2} \gamma \delta \vec{E} - \frac{en_{cr}^2}{m_in_i}\partial_y\Big(\frac{\vec{v}_{cr}}{c}\times \delta \vec{B}\Big)_y
\Big(\frac{\vec{v}_{cr}}{\gamma^2 + k^2c_{s,i}^2}\Big).
\label{eq:curr}
\end{equation}
Thus, combining the Ampere's and Faraday's laws with Eq. (\ref{eq:curr}) we recover Eq. (\ref{eq:dispersion}). 

\section{Appendix B:  The effect of CR charge compensation on the growth of the PCDI}
\label{app:B}
\begin{figure}
  \centering
\includegraphics[width=9cm]{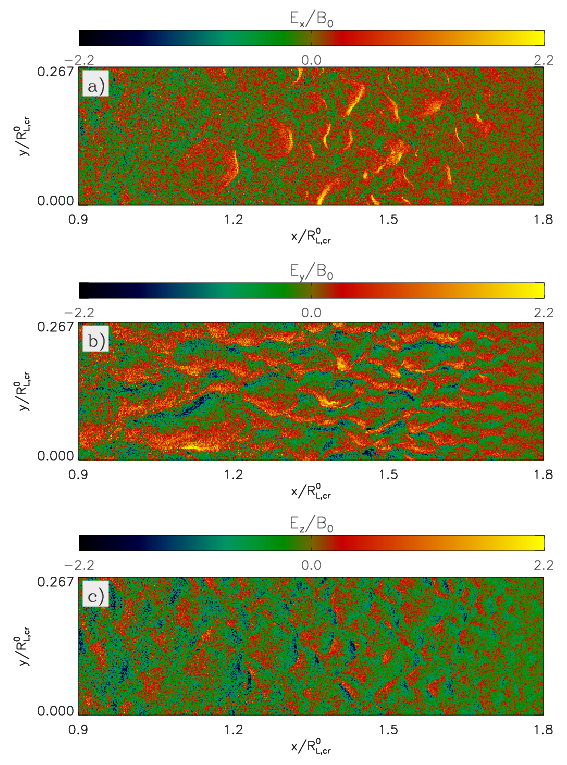}
\caption{The spatial structure of the electric field produced by the PCDI is depicted for a region of simulation $S_*$ at $t\gamma_{PCDI}=32$. The electric field is normalized in terms of $B_0$.}
\label{fig:protruelec}
\end{figure} 
If CRs are made of only positively charged particles, then the background plasma is expected to carry a negative, compensating charge that nearly cancels it. In \S \ref{sec:constcurr} we saw that this background charge may suppress the non-linear growth of the PCDI. In regions where the PCDI produces a decrease in the plasma density, a net positive charge will grow. In the linear regime, this net charge growth is not important. It creates an electric field proportional to the product of the charge fluctuations in the background plasma, which is approximately equal to $\rho_{c,cr}\delta \rho/\rho$, times the characteristic length scale of the fluctuations, which is proportional to $\delta B/B_0$. Thus, the force on the plasma due to this effect is of second order in perturbation theory, and can be neglected. In the non-linear regime, however, this electrostatic force can be relevant. As the PCDI magnetic loops expands, the electric field in the loops will be given by $\vec{E} \sim 4\pi n_{cr} e\vec{r}$, where $\vec{r}$ points in the radial direction. Thus, the loops will feel an electrostatic force $\sim -4\pi \vec{r}n_{cr}^2e^2$, that will push them back to their center. Recalling that, when the amplified field becomes non-linear, $|\vec{r}(t)| \approx \lambda_{PCDI} \approx cB/J_{cr}$, then this force will be of the order of $\vec{J}_{cr} \times \vec{B}$ if the CRs move at an average velocity close to $c$. This implies that the force driving the expansion ($-\vec{J}_{cr} \times \vec{B}/c$) would be roughly compensated by this electrostatic force, which would quench the non-linear growth of the PCDI.

This situation, however, would not happen when the PCDI is driven by CRs protruding into regions of pre-amplified field, as suggested by the simulations presented in \S \ref{sec:crback}. The reason has to do with the way the compensating charge is distributed in the loops. As CRs penetrate into the pre-amplified field regions, their electric current  induces a return current in the plasma ($\vec{J}_{ret} \approx \vec{J}_{cr}$), which is what compensates the charge. However, if this happens while the loops are being formed, this return current will tend to separate charges within the loops, producing a dipolar charge distribution in them.

This can be seen from the shape of the electric field, $\vec{E}$, shown in Figure \ref{fig:protruelec}, which depicts the same simulation region as Figure \ref{fig:protruden}. In the magnetic loops locations, $\vec{E}$ does not have the monopolar shape needed to stop the loop expansion. Instead, we see that, whereas the $y$ component has a monopolar shape, the $x$ component looks like a dipolar field, indicating the presence of charge separation within the loops. In the situation depicted in Figure \ref{fig:protruelec}, electrons accumulate to the right of the loops. Thus, while the negative charge of the plasma concentrates mainly in regions where $\vec{B} || \hat{y}$, the places where $\vec{B} \perp \hat{y}$ will be more neutral. Thus, there will be parts of the loops where the electrostatic forces are less important than in the cases where the charge in the loops is distributed homogeneously (like in the simulations presented in \S \ref{sec:constcurr}), and their evolution will be dominated by the $-\vec{J}_{cr}\times \vec{B}$ force that expands the loops. This way, the PCDI can grow to non-linear amplitudes, even when the CRs may be mainly composed of positively charged particles.


\begin{thebibliography}{}
\bibitem[Axford et al.(1977)]{AxfordEtAl77} Axford, W. I., Leer, E., \& Skadron, G. 1977, 15th Int. Cosmic Ray Conf., 11, 132
\bibitem[Ballet(2006)]{Ballet06} Ballet, J. 2006, Adv. in Space Res., 37, 1902
\bibitem[Bell(1978)]{Bell78} Bell, A. R. 1978, \mnras, 182, 147
\bibitem[Bell(2004)]{Bell04} Bell, A. R. 2004, \mnras, 353, 550
\bibitem[Bell(2005)]{Bell05} Bell, A. R. 2005, \mnras, 358, 181
\bibitem[Blandford \& Ostriker(1978)]{BlandfordEtAl78} Blandford, R. D., \& Ostriker, J. P. 1978, \apj, 221, L29
\bibitem[Buneman(1993)]{Buneman93} Buneman, O. 1993, ``Computer Space Plasma Physics'', Terra Scientific, Tokyo, 67
\bibitem[Giacalone et al.(2007)]{GiacaloneEtAl07} Giacalone, J. \& Jokipii, J.R. 2007, \apj, 663, 41
\bibitem[Krymsky(1977)]{Krymsky77} Krymsky, G. F. 1977, Sov. Phys. Dokl., 23, 327
\bibitem[Kulsrud \& Pearce(1969)]{KulsrudEtAl69} Kulsrud, R., \& Pearce, W. P. 1969, \apj, 156, 445
\bibitem[Lyutikov (2009)]{Lyutikov09} Lyutikov, M. 2009, arXiv:0912.1784
\bibitem[Milosavljevi\'{c} \& Nakar(2006)]{MilosavljevicEtAl06} Milosavljevic, M., \& Nakar, E. 2006, \apj, 651, 979
\bibitem[Niemiec et al.(2008)]{NiemiecEtAl08} Niemiec, J., Pohl, M., Stroman, T., \& Nishikawa, K. 2008, \apj, 684, 1189
\bibitem[Riquelme \& Spitkovsky(2009)]{RiquelmeEtAl09} Riquelme, M. A., \& Spitkovsky, A. 2009, \apj, 694, 626
\bibitem[Sironi \& Goodman(2007)]{SironiEtAl07} Sironi, L., \& Goodman, J. 2007, \apj, 671, 1858
\bibitem[Spitkovsky(2005)]{Spitkovsky05} Spitkovsky, A. 2005, AIP Conf. Proc, 801, 345, astro-ph/0603211
\bibitem[Stroman et al.(2009)]{StromanEtAl09} Stroman, T., Pohl, M., Niemiec, J. 2009, \apj, 706, 38
\bibitem[Uchiyama et al.(2007)]{UchiyamaEtAl07} Uchiyama, Y., Aharonian, F. A., Tanaka, T.,  Takahashi, T., \& Maeda, T. 2007, Nature, 449
\bibitem[V\"{o}lk et al.(2005)]{VolkEtAl05} V\"{o}lk, H. J., Berezhko, E. G., \& Ksenofontov, L. T. 2005, \aap, 433, 229
\bibitem[Zirakashvili et al.(2008)]{ZirakashviliEtAl08} Zirakashvili, V. N., Ptuskin, V. S., \& Volk, H. J. 2008, \apj, 678, 255
\end{thebibliography}
\end{document}